# High performance fiber-coupled NbTiN superconducting nanowire single photon detectors with Gifford-McMahon cryocooler

Shigehito Miki, Taro Yamashita, Hirotaka Terai and Zhen Wang

Advanced ICT Research Institute, National Institute of Information and Communications Technology, 588-2, Iwaoka, Nishi-ku, Kobe, Hyogo 651-2492, Japan

**Abstract**

We present high performance fiber-coupled niobium titanium nitride superconducting nanowire single photon detectors fabricated on thermally oxidized silicon substrates. The best device showed a system detection efficiency (DE) of 74%, dark count rate of 100 c/s, and full width at half maximum timing jitter of 68 ps under a bias current of 18.0 μA with a practical Gifford-McMahon cryocooler system. We also introduced six detectors into the cryocooler and confirmed that the system DE of all detectors was higher than 67% at the dark count rate of 100 c/s.

## 1. Introduction

Superconducting nanowire single-photon detectors (SNSPDs or SSPDs [1]) have emerged as promising photon detectors that can be applied in optical quantum information and communication technology, fundamental quantum optics studies, classical laser communication technology, and light detection and ranging technology. The SSPDs potentially have broadband sensitivity that ranges from the visible to the near-infrared wavelengths, high detection efficiency (DE), low timing jitter, high count rate, and low dark count rate (DCR) [2]. They are already in use in many types of applications, such as the ones described above, and their superiority among detectors have been demonstrated [3-6]. Further, significant effort has been expended to further improve the system performance of these detectors. In particular, a primary objective is to identify a method to simultaneously achieve high system DE, low DCR, and low timing jitter in a practically available SSPD system, which is expected to significantly influence various SSPDs applications.

Recently, two remarkable studies reported the achievement of high system DE, low DCR, and low timing jitter [7,8]. F. Marsili et al. reported that WSi SSPDs achieved extremely high system DE of ~93% due to the low energy gap of its material [7]. This value certainly makes the WSi SSPDs attractive for use in various applications. However, the $T_c$ of these films is extremely low to operate in Gifford-McMahon (GM) cryocooler systems while retaining their performances. D. Rosenberg et al. studied four-pixel NbN SSPD array on thermally oxidized Si substrates installed into GM cryocooler system, and reported high system DE of 76%, low DCR of ~100 c/s, and timing jitter of

80 ps [8]. The authors also succeeded in showing the potential of their method to realize large-scale SSPD arrays. However, they used four-pixel array systems with a low temperature nanopositioner, which makes it highly complicated to realize independent, multichannel systems, required for specific applications such as quantum key distribution experiments. In addition, in both reports, the switching current $I_{sw}$ was relatively small. The operation of SSPDs with a small $I_{sw}$ is supposed to be unstable because of the contribution of the external environment (grounding noise, experimental setup connected to SSPD system, and more) to the fluctuation of the bias current's flow to the devices cannot be ignored. These detectors also require a cryogenic amplifier to achieve short timing jitter, making the cryocooler system complicated [7,8].

Niobium titanium nitride (NbTiN) can be viewed as an adequate candidate for use in SNSPDs because its $T_c$ is lower than that of NbN films, thus enabling higher sensitivity of the SNSPD system to incident photons [9,10]. The $T_c$ of films is sufficiently higher than 2.3 K, enabling stable operation in GM cryocooler systems. In this paper, we report the fabrication of high performance NbTiN-SSPDs with a practical GM cryocooler system, which is available for use in various applications. We describe the fabrication of SSPD devices based on NbTiN nanowire on thermally oxidized Si substrates, as well as the simulation results of our device's absorptance. Then, we verify the system DE, DCR, and timing jitter of our devices that are installed in a practical multichannel system. Our SSPD system has the following concomitant features that satisfy practical requirements and allow easy access from potential users. First, our detector system can achieve high system DE, low DCR, low timing jitter, and high $I_{sw}$ at the same time. Second, our SSPD system uses fiber-coupled package technology, which is useful to achieve high optical coupling efficiency and install multi-packaged devices in a cryocooler system. Third, our system is based on compact GM cryocooler system with an operation temperature of 2.3K, which allows turnkey and continuous operation with low power consumption.

## 2. Device design and fabrication

Prior to device fabrication, both surfaces of the Si substrate were thermally oxidized to 270 nm thick $SiO_2$ layers. The $SiO_2$ layer on the reserve side of the substrate functioned as an anti-reflection layer for 1550 nm wavelength and the layer at the front side functioned as a component of cavity structure. The NbTiN thin films that were used for the nanowire were deposited by DC reactive sputtering at an ambient substrate temperature [11]. We deposited 5 nm thick NbTiN films. The film thickness was controlled by the deposition rate and deposition time. The NbTiN films were formed into a 100 nm wide and 60(100) nm spaced meandering nanowire that covered a square area of 15 μm × 15 μm. A 250 nm thick SiO and a 100 nm thick Ag mirror covered the nanowire to enhance the absorption efficiency, $P_{absorb}$, of photons to the meandering area of the nanowire. The thickness of both the $SiO_2$ and SiO layers was chosen to be λ/4 of 1550 nm.

Fig. 1 shows the schematic presentation of the fabricated NbTiN-SSPD device. The structure of the cavity in this work was almost the same as the one that was reported in [8] as an effective structure to enhance the absorption $P_{absorb}$ of photons that are incident on the nanowire. To verify the actual effectiveness of this structure, we simulated the $P_{absorb}$ of this cavity structure and compared the results to a simple cavity structure that was reported in [12]. For the simulation of $P_{absorb}$, we used the software that enables modeling of the electromagnetic behavior of single-period multilayer gratings (PhotonicsSHA-2D) [13]. In this software, the incident wave enters the unit cell corresponding to the device with TE/TM polarization, and transmission as well as reflection can be calculated for an arbitrary wavelength range. All of the designs that were used in the simulation were those described above. In the simulation, we used complex refractive indices as $n_{Si}$ = 3.63, $n_{SiO2}$ = 1.44, $n_{SiO}$ = 1.551, $n_{NbTiN}$ = 4.563 + i4.911, $n_{Ag}$ = 0.322 + i10.99. These values were measured beforehand using a spectroscopic ellipsometer, and constant values at 1550 nm wavelength were used. For the simplicity of the simulation, we assumed the incident light to be irradiated from the Si medium, and hence, light reflection from the back side of the Si substrate was not considered in the simulation. Fig. 2 shows the simulated $P_{absorb}$ of the devices coupled with TE mode light as a function of wavelength. For comparison, this same Fig. also shows the simulated results of $P_{abosrb}$ for SSPD on MgO substrate ($n_{MgO}$ = 1.720) with same design except the substrate. Results of the simulation show that, for wavelengths ranging from 1400 to 2000 nm, the $P_{absorb}$ is clearly higher than that of simple cavity structure, and a high value of 97% is observed at the wavelength of 1550 nm.

### 3. Cryocooler system and experimental setup

In our previously published work, we developed a practical SNSPD system based on two-stage-type Gifford-McMahon (GM) cryocoolers [12,14]. Here, we briefly introduce our cryocooler systems and experimental setup. Fiber-coupled compact packages that utilize fiber-spliced graded index (GRIN) lenses were used to achieve efficient optical coupling between incident photons and device sensitive area [12]. The packaged devices were then installed into the GM cryocooler system with a 1.5 kW rated input power consumption at a driving frequency of 60 Hz. This cryocooler system can cool the sample stage to 2.3 K, within thermal fluctuation of 10 mK. Further, it can set maximally six SSPD packages for use in various applications that require several independent detector channels, such as quantum key distribution experiments [3-5]. The semi-rigid coaxial cables and single mode (SM) optical fibers for telecommunication wavelength were introduced into each package, and the optical fibers in a cryocooler system made several loops with a diameter of ~30 mm to mitigate the dark counts that originated from the blackbody radiation [15]. The semi-rigid coaxial cables were connected to a bias tee and two room temperature low noise amplifiers at the outer side of the cryocooler.

For the measurements of system DE, a continuous tunable laser was used as an input photon source and they were heavily attenuated so that the photon flux at the input connector of the cryostat was $10^6$ photons/s which is sufficiently low to keep the linearity of the output counts for the incident photon flux. A fiber polarization controller was inserted in front of the cryocooler's optical input in order to control the polarization properties of the incident photons so as to maximize the DE. The wavelength was tuned to maximize the DE in the range from 1540 to 1560 nm, to avoid the influence of periodic fringes resulting from the interference among the multi-optical layer boundaries [16]. Although the optical losses from the input port to the device in our GM cryocooler system were confirmed to be 0.2–0.4 dB, this effect was not considered in the calibration of input photon flux. It is noted that these losses were mainly due to the fiber connector that was placed in the cryocooler system to simplify the detector exchanges, and could be easily improved by splicing the fibers. Consequently, the system DE was obtained by subtracting DCR from the output current rate and by dividing this value by the value of input photon flux.

The timing-jitter characteristics were measured using a time-correlated single-photon counting module with a 1 ps resolution (HydraHarp 400, PicoQuant GmbH). A 1550 nm wavelength pulsed laser with a 100 fs pulsewidth was attenuated to be lower than 0.3 photon/pulse and used as the photon source. The temporal correlations between the synchronized trigger pulses from pulsed laser module and the output pulses from the device through two room temperature low-noise amplifiers were recorded with timing module.

## 4. Performance in the practical GM cryocooler system

Fig. 3(a) shows the system DEs, DCR and full width at half maximum (FWHM) timing jitter of fabricated SSPD device (channel 1 in Table 1) with 100 nm wide and 60 nm spaced nanowire, as a function of bias current installed in GM cryocooler system. The $I_{sw}$ of a device was 19.2 μA, which was sufficient to achieve stable operation and relatively short timing jitter. Although the system DE reached maximal values of 77.3% when the bias current was introduced near $I_{sw}$, this bias condition may not be suitable for practical use because the DCR increased to over $10^4$ c/s. In the meantime, the device could achieve high system DE, low DCR, and short timing jitter simultaneously even for lower values of bias current. For example, the device showed a system DE of 74.0%, DCR of 100 c/s, and timing jitter of 68 ps at a bias current of 18.0 μA, which makes it considerably attractive for use in various applications. Thus far, we have observed a significant improvement in system DE and DCR by lowering the operation temperature [15]. Thus, we also measured the system DE and DCR under the operation temperature of 0.3 K, using the $^3$He cryocooler system; results of these measurements are shown in Fig. 3(b). These results disclose that the device does not show the significant improvement of system DE when operation temperature is changed; this can be explained by the following reasoning. Since the system DEs clearly showed sigmoid dependence on the bias current as reported in [17], we can use the method of least squares to estimate the numerical values of the pulse generation probability after photon absorption to the nanowire $P_{pulse}$. Accordingly, the $P_{pulse}$ was estimated to be 94.1% even at 2.3 K, when the current was biased near $I_{sw}$. Although, as shown in the inset of Fig. 3(b), the $P_{pulse}$ reached the saturated region in the experiment with lowered temperature, the system DE remained at 82.7%.

The dependence of the bias current on the dark counts has two different slopes, which are due to two different effects, as reported in [15]. For bias currents higher than ~17.5 µA, the dark counts are predominantly caused by intrinsic dark counts that originate from vortex-antivortex unbinding [18]. To avoid these intrinsic dark counts, the dependence of the bias current on system DE must have saturated region so as to achieve high system DE for low bias current levels, at which the intrinsic dark counts are not present. Although decreasing the operation temperature is simple and effective way to realize, a relatively large cooling system is necessary. Thus, it is preferable to achieve improvement in a way that does not require decreasing the operation temperature, for example, by optimizing superconducting characteristics and nanowire designs. On the other hand, the dark counts for bias currents lower than 17.5 µA are caused by the blackbody radiation. We confirmed that the DCR can be reduced less than ~1/10 without decreasing the system DE, by making the fiber loops in a cryocooler. Insertion of effective band path filter would be necessary to further reduce DCR by the blackbody radiation.

The timing jitters were observed with two conventional readout configurations: one had a 50 Ω resistor connected in parallel to the device, and the other had no shunting resistor. In addition, we used commercial room temperature low noise amplifiers (Chain of LNA-550 and LNA-1000, RF Bay Inc.), which must have poorer performance than cryogenic amplifiers. Nevertheless, because of the device's high $I_{sw}$, we could achieve timing jitters that were lower than those reported in [7,8]. We observed that the shortest timing jitter of 51 ps occurred at a bias current of 16 µA without a shunting resistor. However, for higher values of bias current without the shunting resistor, the device latched into the normal state and stopped functioning. Although the 50 Ω shunting resistor is, thus, indispensable to achieve high system DEs in this conventional readout configuration, new readout electronics aiming to avoid the latching in the high bias-current region without the 50 Ω shunting resistor would be effective. In fact, we have achieved a short timing jitter of 37 ps at a bias current value of 18 µA for the device with similar $I_{sw}$, using single-flux-quantum readout circuit without latching near $I_{sw}$ [19].

For the actual applications, we have built an SSPD system with six independent channels, and Table 1 lists the superconducting and optical characteristics of each channel. In the process of building up the system, the devices with 100 nm width and 100(60) nm spacing were used. Although the SSPDs with 100 nm width and 60 nm spacing showed slightly higher values than others, the system DEs of all channels were higher than 67% at the DCR of 100 c/s. Since $I_{sw}$'s of all devices showed similar values and were centered at ~20 µA, the timing jitter values are expected to be comparably short, similar to those shown in Fig. 3(a).

## 5. Conclusions

We have verified the performance of NbTiN SSPDs fabricated on a thermally oxidized Si substrate installed into compact fiber-coupled packages. We successfully demonstrated that the fiber-coupled packaged device, installed in a practical GM cryocooler system, simultaneously showed a system DE higher than 74% at a wavelength of 1550 nm, low DCR of 100 c/s, and short timing jitter of 68 ps, at the large bias current value of 18.0 µA; these values are significantly higher than those reported in our previous works [12]. We also succeeded in building a practical SSPD system with 6 independent channels, and all channels showed system DE values higher than 67% at the 100 c/s DCR, which can be instantly utilized for various applications. Ongoing studies aim to further improve the system DE, DCR, and timing jitter, for example, by improvement of $P_{absorb}$, insertion of effective filter to cut the blackbody radiation, and improvement of the readout circuit to avoid latching behavior.

## Acknowledgements

The authors thank Saburo Imamura and Makoto Soutome of the National Institute of Communications Technology for their technical support. The authors also thank Mikio Fujiwara and Masahide Sasaki of the National Institute of Communications Technology for the fruitful discussions.

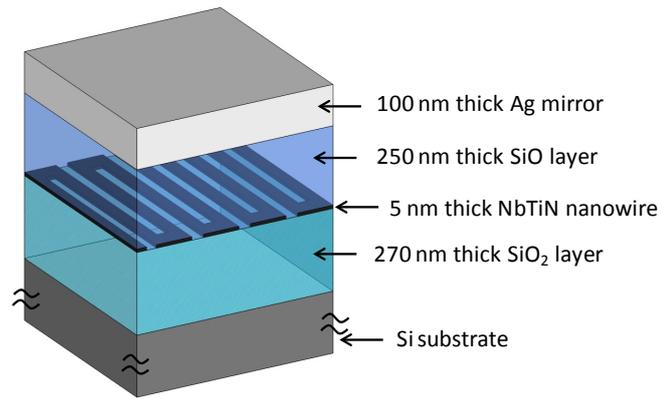

Fig. 1 Schematic configuration of NbTiN SSPD on thermally oxidized Si substrate.

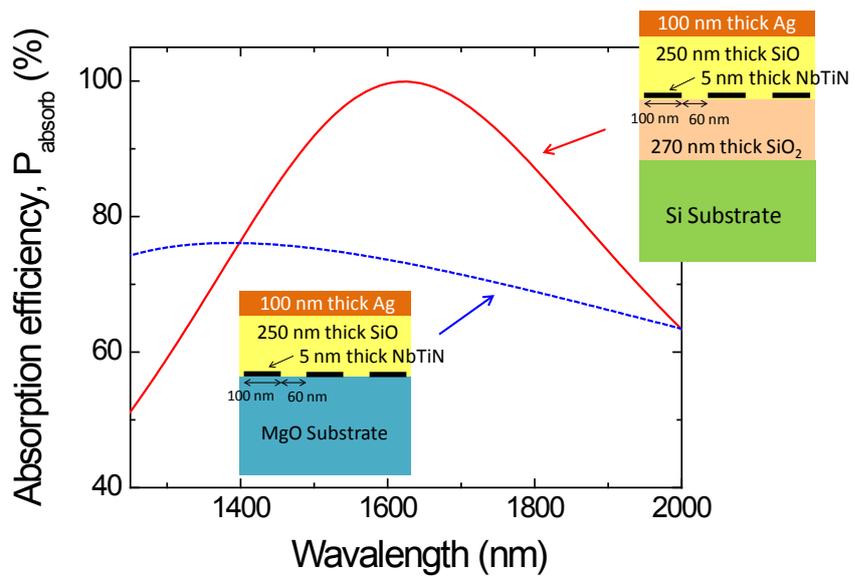

Fig. 2 Simulated $P_{absorb}$ of NbTiN SSPD devices on thermally oxidized Si substrate. For comparison, $P_{absorb}$ of NbTiN SSPD on MgO substrate with simple cavity structure is also shown in this Fig..

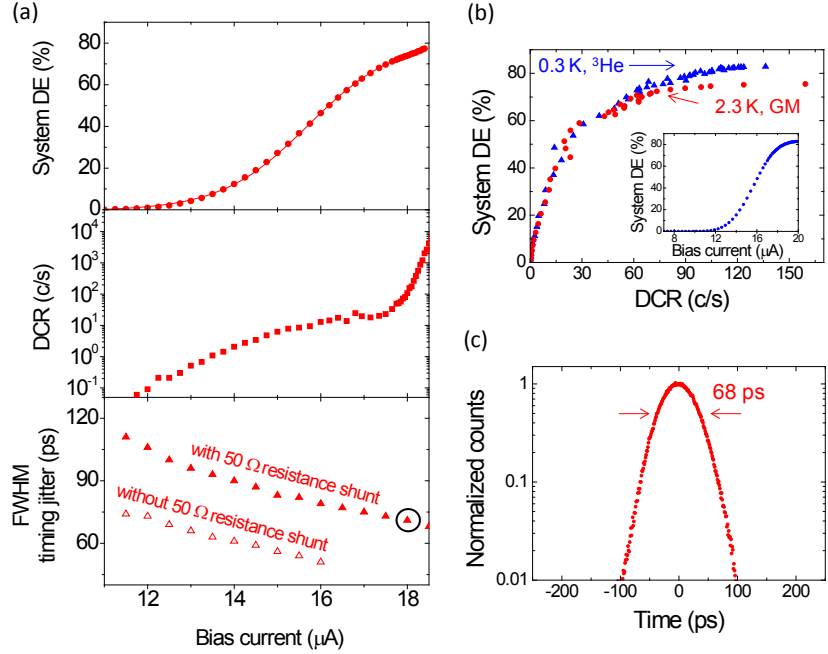

Fig. 3 (a) System DE, DCR, and FWHM timing jitter as a function of bias current for fabricated device with 100 nm wide and 60 nm spaced meandering installed into GM cryocooler system. The solid line in system DE dependencies is fitted by the sigmoid function. (b) System DE as a function of DCR, for different operation temperatures obtained by using the GM and $^3$He cryocooler systems, respectively. Inset shows the system DE as a function of bias current at the operation temperature of 0.3 K. (c) Histogram of timing jitter obtained with GM cryocooler system at a bias current of 18.0 μA with a 50 Ω shunting resistor (circled point in Fig.3(a)).

Table 1. Nanowire design, superconducting, and optical characteristics of six-channel SSPD system. The parameter $P_{sys,100DCR}$ is the system DE at the DCR of 100 c/s, and the parameter $P_{sys,max}$ is the maximal system DE for bias currents near $I_{sw}$ with higher DCR than ~1,000 c/s..

| Channel | width/space (nm) | $T_c$ (K) | $I_{sw}$ (μA) | $P_{sys,100DCR}$ (%) | $P_{sys,max}$ (%) |
|---|---|---|---|---|---|
| 1 | 100/60 | 7.54 | 19.2 | 74.0 | 77.3 |
| 2 | 100/60 | 7.49 | 20.6 | 73.0 | 76.6 |
| 3 | 100/60 | 7.52 | 20.0 | 67.9 | 75.6 |
| 4 | 100/100 | 7.48 | 18.2 | 67.0 | 71.7 |
| 5 | 100/100 | 7.52 | 20.2 | 67.5 | 72.7 |
| 6 | 100/100 | 7.52 | 19.8 | 67.3 | 71.0 |